\documentclass[10pt,conference]{IEEEtran}
\IEEEoverridecommandlockouts
\usepackage{cite}
\usepackage{amsmath,amssymb,amsfonts}
\usepackage{algorithmic}
\usepackage{graphicx}
\usepackage{textcomp}
\usepackage{xcolor}
\usepackage{glossaries}
\usepackage{tcolorbox} 
\usepackage{multirow,booktabs}
\usepackage{listings}
\usepackage{enumitem}
\usepackage{url}

\def\BibTeX{{\rm B\kern-.05em{\sc i\kern-.025em b}\kern-.08em
    T\kern-.1667em\lower.7ex\hbox{E}\kern-.125emX}}

\begin{document}

\title{Your Build Scripts Stink:\\ The State of Code Smells in Build Scripts}

\makeatletter
\newcommand{\linebreakand}{%
  \end{@IEEEauthorhalign}
  \hfill\mbox{}\par
  \mbox{}\hfill\begin{@IEEEauthorhalign}
}
\makeatother

\author{%
\IEEEauthorblockN{Mahzabin Tamanna}
\IEEEauthorblockA{%
    \textit{North Carolina State University}\\
    North Carolina, USA\\
    mtamann@ncsu.edu\\
}
\and
\IEEEauthorblockN{Yash Chandrani}
\IEEEauthorblockA{%
    \textit{North Carolina State University}\\
    North Carolina, USA\\
    ychandr@ncsu.edu\\
}
\and
\IEEEauthorblockN{Matthew Burrows}
\IEEEauthorblockA{%
    \textit{North Carolina State University}\\
    North Carolina, USA\\
    myburrow@ncsu.edu
}
\linebreakand 
\IEEEauthorblockN{Brandon Wroblewski}
\IEEEauthorblockA{%
    \textit{North Carolina State University}\\
    North Carolina, USA\\
    bnwroble@ncsu.edu
}
\and
\IEEEauthorblockN{Laurie Williams}
\IEEEauthorblockA{%
    \textit{North Carolina State University}\\
    North Carolina, USA\\
    lawilli3@ncsu.edu
}
\and
\IEEEauthorblockN{Dominik Wermke}
\IEEEauthorblockA{%
    \textit{North Carolina State University}\\
    North Carolina, USA\\
    dwermke@ncsu.edu
}
}

\maketitle

\begin{abstract}
Build scripts automate the process of compiling source code, managing dependencies, running tests, and packaging software into deployable artifacts.
These scripts are ubiquitous in modern software development pipelines for streamlining testing and delivery. While developing build scripts, practitioners may inadvertently introduce code smells, which are recurring patterns of poor coding practices that may lead to build failures or increase risk and technical debt. \textit{The goal of this study is to aid practitioners in avoiding
code smells in build scripts through an empirical study of
build scripts and issues on GitHub.}
We employed a mixed-methods approach, combining qualitative and quantitative analysis. First, we conducted a qualitative analysis of 2000 build-script-related GitHub issues to understand recurring smells. Next, we developed a static
analysis tool, \textit{Sniffer}, to automatically detect code smells in 5882 build scripts of Maven, Gradle, CMake, and Make files, collected from 4877 open-source GitHub repositories. To assess \textit{Sniffer's} performance, we conducted a user study, where Sniffer achieved higher precision, recall, and F-score. We identified 13 code smell categories, with a total of 10,895 smell occurrences, where 3184 were in Maven, 1214 in Gradle, 337 in CMake, and 6160 in Makefiles.

Our analysis revealed that \textit{Insecure URLs} were the most prevalent code smell in Maven build scripts, while \textit{Hardcoded Paths/URLs} were commonly observed in both Gradle and CMake scripts.
\textit{Wildcard Usage} emerged as the most frequent smell in Makefiles. The co-occurrence analysis revealed strong associations between specific smell pairs of \textit{Hardcoded Paths/URLs} with \textit{Duplicates}, and \textit{Inconsistent Dependency Management with Empty} or \textit{Incomplete Tags}, which indicate potential underlying issues in the build script structure and maintenance practices. Based on our findings, we also recommended strategies to remove code smells in build scripts to improve the efficiency, reliability, and maintainability of software projects.

\end{abstract}
\maketitle

\section{Introduction}\label{intro}
Build scripts are ubiquitous in modern software development, facilitating the automation of complex software application compilation, testing, and deployment processes.
Build scripts are used to transform source code into deliverable artifacts.
Developers use build tools such as Maven and Gradle to define their respective build systems~\cite{desarmeaux2016dispersion} and maintain consistency and reproducibility across diverse environments by standardizing the build process~\cite{dev,DeployHQ,LaunchDarkly}. To automate the software development and testing process, the majority of well-maintained software projects make use of build tools like Maven and Gradle~\cite{stackover2025}. According to JetBrains~\cite{jetbrain}, Maven and Gradle are popular for project building with 71\% and 48\% usage in the industry, respectively.

As software systems grow in scale and functionality, build scripts become increasingly complex~\cite{mcintosh2011empirical} and might require frequent maintenance~\cite{hassan2017change}, which can negatively impact overall project quality and developer productivity~\cite{mcintosh2012evolution}. Consequently, these scripts become susceptible to various code smells.  

Code smells are indicators of potential underlying issues in the design and implementation of software, which increases the fault-proneness~\cite{khomh2009exploratory} and makes the system more difficult to maintain~\cite{sjoberg2012quantifying} and understand~\cite{sahin2014code}. Code smells are the symptoms of poor design and implementation~\cite{fowler2018refactoring}, which leads to technical debt~\cite{shull2013technical}. Code smell introduces bugs, affects software maintainability~\cite{yamashita2012code}, and causes developers' burden~\cite{yamashita2013developers}. Early detection and remediation of these smells could aid in reducing higher maintenance costs, improving software reliability, and mitigating security risks throughout the automated software development life cycle~\cite{cordeiro2024empirical}.

Prior studies showed that files without code smells exhibit approximately 33\%--65\% lower risk of fault compared to those containing code smells~\cite{johannes2019large,Saboury2017}.
Code smells are important predictors of build failures, indicating a link between code quality in build scripts and build reliability~\cite{barrak2021builds}.
Moreover, broken builds could disrupt team productivity and negatively impact overall project performance~\cite{paixao2017interplay}.
Previous studies have explored the adverse effects of code smells in platforms such as Travis CI~\cite{saidani2021toward}, Infrastructure as Code (IaC)~\cite {rahman2019seven}, and source code~\cite{di2018detecting}.
However, despite the widespread use of build scripts in modern software development, particularly within Continuous Integration and Continuous Deployment (CI/CD) pipelines, to our knowledge, a systematic analysis of the presence of code smells in the context of build scripts has not been performed yet.
Overlooked code smells in build scripts could inadvertently facilitate the spread of bad coding practices, thereby impeding the adaptability and maintainability of software systems~\cite{arcoverde2011understanding,lozano2007assessing}.
In this paper, we aim to systematically investigate, identify, and quantify the prevalence of code smells in the context of build scripts, employing both static analysis techniques and qualitative assessments.


\begin{center}
    \textit{The goal of this paper is to aid practitioners in avoiding code smells in build scripts through an empirical study of build scripts and issues on GitHub.}
\end{center}

\noindent
For this, we aim to address the following research questions:

\begin{itemize}[leftmargin=*,topsep=.2em]
    \item[]\textbf{RQ1:} What code smells occur in build scripts?
    \item[] \textbf{RQ2:} How frequently do code smells occur in build scripts?
\end{itemize}
\vspace{.2em}

\noindent
We address our research questions by examining four types of build scripts: Maven~\cite{maven}, Gradle~\cite{gradle}, CMake~\cite{cmake}, and Make~\cite{make}, as well as build script-related GitHub issues, which are collected from open-source GitHub repositories.
We conducted a qualitative analysis of 2,000 GitHub issues to identify prevalent code smells, following established guidelines for qualitative research~\cite{tenny2017qualitative}.
Next, we collected build scripts from GitHub and leveraged Large Language Models (LLMs) to identify potential code smells in build scripts.
We used six general and code-oriented LLMs, i.e., ChatGPT4, ChatGPT4o, Codellama, Llama 8B, Llama 13B, and Mistral, to detect potential code smells in the build script. Through the analysis, we identified a code smell taxonomy and identified 13 code smell categories. Furthermore, we mapped each code smell category to its corresponding Common
Weakness Enumeration (CWE) to establish standardized categorizations of the underlying weaknesses.
Based on our qualitative study and findings from the LLMs, we developed a static analysis tool, \textit{``Sniffer''}, to automatically detect code smells.
Further, we developed an Oracle through a user study to compare our tool's performance against human evaluation.
For our empirical analysis, we gathered 5882 build scripts from 4877 open-source repositories. We measured the occurrence, frequency, smell density, and co-occurrence of the code smells.
The contributions of our study are outlined below:

\begin{enumerate}[leftmargin=*,topsep=.2em]
    \item Identification and detailed classification of code smells specifically prevalent in build scripts for Maven, Gradle, CMake, and Make. 
    \item Development and evaluation of "Sniffer," a novel static analysis tool to automatically detect code smells in build scripts.
    \item Development of an Oracle to compare our linter performance against the human evaluation of code smells.
    \item  Analysis of the occurrence and frequency of build script-specific code smells.
\end{enumerate}


We organized this paper as follows: background information with related work in 
Section~\ref{background}, the methodologies and results of identifying code smells in Section~\ref{code_smell_identify}.
Development methodology and evaluation of Sniffer in Section~\ref{sniffer_development}.
Empirical Analysis of Build Scripts in Section~\ref{em_analysis}.
Detailed discussion and recommendation in Section~\ref{discussion}.
Sections~\ref{ethics} and~\ref{thtb} cover threats to validity. Section~\ref{conclusion} provides conclusions.


\section{Background and Literature Review}\label{background}
The concept of code smells was popularized by Kent Beck on WardsWiki in the late 1990s, and later usage of the term increased by Fowler~\cite{fowler2018refactoring}, which was initially focused on identifying patterns in source code that may indicate deeper design or maintenance problems.
Later, multiple works focused on detecting code smells in various domains, including source code~\cite{liu2019deep,kovavcevic2022automatic, olbrich2010all}, configuration code~\cite{sharma2016does}, and  Infrastructure as Code (IAC)~\cite{rahman2019systematic}.

Code smells act as strong indicators of broader software quality issues.
Barrak et al.~\cite{barrak2021builds} demonstrate that code and test smells are strong predictors of build failures.
According to Adams et al.~\cite{adams2008evolution}, there is a co-evolutionary relationship between source code and build scripts, emphasizing that both need to remain synchronized. Build scripts are prevalently used in automating the compilation, testing, packaging, and deployment of software and are also susceptible to similar code smells.
By encapsulating a sequence of commands and configurations, build scripts eliminate the need for manual intervention, reducing the risk of human error and accelerating the software delivery pipeline~\cite{kawalerowicz2013classification}. However, when smells in build scripts are neglected, they often lead to build failures~\cite{hassan2018hirebuild}. 
The effort of maintaining build systems is nontrivial.
Research by McIntosh et al.~\cite{mcintosh2011empirical} found that build maintenance tasks account for up to 27\% of the overhead in source code development and 44\% in test development.
Moreover, 22\% of commits and 27\% of development tasks directly involve build scripts, highlighting the substantial developer effort spent on managing build infrastructure. Code smells can lead to broken builds and integration delays, disrupting team productivity and software delivery timelines~\cite{vassallo2019automated}. Zhang et al.~\cite{zhang2022buildsonic} analyzed CI performance smells, reporting a 12.4\% improvement in build performance when these smells were addressed. Similarly, Saidani and Ouni~\cite{saidani2021toward} demonstrated that incorporating bad smell detection enhances build failure prediction accuracy by 4\%, reinforcing the value of systematic smell detection.

Despite the importance of build scripts in modern software development workflows, build scripts remain understudied. While source code smells have been extensively studied and cataloged in other codebases~\cite {liu2019deep,palomba2013detecting,rasool2015review,santos2018systematic,fawad2025android}, there is no comprehensive or standardized taxonomy of smells for build scripts that has been studied.
Moreover, security incidents such as the XZ Utils backdoor~\cite{XZ} have underscored how weaknesses and negligence in build processes can act as a vector for supply chain attacks.
These highlight the gap and need for a systematic analysis of build script quality to identify both maintenance and security-related smells.
To address this gap, this work aims to systematically identify and categorize code smells specific to build scripts across multiple build systems and to suggest their potential mitigation to enhance the maintainability and security of build automation processes.



\section{Investigating Code Smells in Build Scripts}\label{code_smell_identify}

We conducted a mixed-method study with qualitative and quantitative parts to answer our research questions. We discussed the methods that have been applied for RQ1 in Section~\ref{RQ1-meth} and the findings for RQ1 in Section~\ref{RQ1-res}.

\subsection{RQ1: What code smells occur in build scripts?}~\label{RQ1-meth}
In steps of methodology: i) Qualitative analysis of GitHub issues. ii) Qualitative analysis of build scripts LLMs. iii) Code smells to CWE mapping.
\subsubsection{\textbf{Qualitative Analysis on GitHub Issues}} Our qualitative analysis consists of three phases: platform selection, data collection, and qualitative coding.
\\
\textit{\textbf{Platform Selection:}} For this study, we selected GitHub as it is a widely used platform for open source software development~\cite{ITpro} and scientific research~\cite{cosentino2017systematic}. According to the StackOverflow 2024 survey~\cite{stackoverflow}, Maven, Gradle, Cmake, and Make are the most popular and widely used build system tools; as such, we selected these build tools for our study.\\
\textit{\textbf{Data Collection:}} We performed a keyword-based search within the GitHub repositories using the Rest API and GraphQL. We specifically searched for build scripts and code-smell-related keywords in the GitHub repositories and issues' metadata, including titles, descriptions, and labels for the selected build type. Two authors collaboratively reviewed and iteratively refined the search queries for their relevance and accuracy based on prior studies on code smells and practitioner blog posts~\cite{pragmatic,sharma2016does,azeem2019machine,codinghorror,jetbraincode}.
The search strings included:\\[.25em]
\noindent\fbox{%
    \parbox{0.96\linewidth}{%
        ``codesmell'', ``code smell'', ``bad practice'', ``bad smell'', ``HTTP external download", ``empty password", ``long parameter", ``long method", ``inconsistent version", ``abandoned dependencies", ``unused dependency", ``duplicate code", ``long method", ``admin by default", ``anti-pattern", ``hardcoded paths", ``duplicated configuration", ``duplicate code", ``conditional complexity'', ``Dead code'', ``inconsistent name'', ``complex build logic", ``inconsistent dependency versions", ``unused dependencies", ``excessive build times", ``lack of modularity", ``excessive task dependencies", ``Path-misconfiguration".
    }%
}\vspace{.25em}
Following the process, we collected 7533 GitHub issues. Further removing the duplicates, 7104 unique issues remained.

\textit{\textbf{Qualitative Coding:}}
We conducted a qualitative analysis to obtain a summarized overview of recurring bad coding practices in build scripts and to gain insight into potential code smells. For this process, we randomly selected 200 issues of each type of build script (Maven, Gradle, CMake, and Make), totaling 800 GitHub issues out of 7104 unique issues. Two authors independently conducted a descriptive qualitative analysis~\cite{saldana2021coding} on these sampled issues. To our knowledge, at the time of this study, no prior research or predefined list of code smells specific to build scripts was available.
To enhance the completeness of our qualitative findings, we compared the 800 issues analyzed with the remaining dataset using the Jaccard similarity index~\cite{niwattanakul2013using}. We employed the Jaccard score to identify and prioritize dissimilar issues for further analysis based on the assumption that similar issues are likely to yield similar or the same code smell as detected in the first 800 issues.
Consequently, issues with the lowest Jaccard scores were selected as starting points for further investigation.
Following the process, we analyzed an additional 1200 issues, 300 issues from each building type, ultimately reaching a saturation point~\cite{guest2006many}, where no new code smells were identified.
Following this process, we qualitatively coded 2000 GitHub issues in total.
After conducting descriptive coding, the authors collaboratively finalized the code smells list and addressed any conflicts through a negotiated agreement approach~\cite{campbell2013coding}.
Disagreements were resolved either by eliminating inappropriate categories for code smells or by merging closely related categories to form a unified and coherent categorization.
Following the processes, we created an initial list of code smells based on our GitHub issue analysis~\cite{datashare}. 

\subsubsection{\textbf{Qualitative Analysis of Build Scripts Leveraging LLMs}}\label{Build_QA}
In this phase, we leveraged Large Language Models (LLMs) to analyze and detect code smells within build scripts.
The primary objective of employing LLMs was to explore their potential for automating the identification of code smells, given their advanced capabilities in understanding both natural language and programming code semantics.
Recent studies demonstrated that LLMs exhibit strong performance in a range of software engineering and security tasks~\cite{chen2021evaluating, williams2024research,zheng2025towards}. Details about the build scripts analysis using LLMs are given below.

\textit{{\textbf{Data Collection:}}}
For our build script analysis, we applied a multi-faceted approach to construct a dataset of build scripts from open-source projects. We developed an automated Python script that utilizes the GitHub API to identify and retrieve build files specifically associated with four selected build systems: Maven, Gradle, CMake, and Make. The script was configured to search for build script filenames corresponding to these systems, such as:
\begin{itemize}
    \item \textbf{Maven:} \texttt{pom.xml}
    \item \textbf{Gradle:} \texttt{build.gradle} and \texttt{build.gradle.kts}
    \item \textbf{CMake:} \texttt{CMakeLists.txt}
    \item \textbf{Make:} \texttt{Makefile} (and files with the \texttt{.mk} extension)
\end{itemize}
For each build system, we employed a system-specific search query (e.g., \texttt{ filename:pom.xml}) and iteratively retrieved the search results from GitHub. To maximize coverage, multiple pages of results were fetched. For search results, we collected relevant metadata, including the repository name, repository URL, and the raw URL of the build script. Next, we collected the exact build script using the corresponding URL. The data collection steps utilized broad GitHub queries and retrieved results from multiple pages, without restrictions on project domain, size, or popularity. This approach allowed us to collect a diverse set of build scripts from projects of varying maturity levels and development practices. Following the data collection step and removing duplicate entries, our final dataset comprised 2,134 distinct build scripts.

\textit{\textbf{Selection of LLM:}} We leveraged LLMs to detect code smells in build scripts. We selected six LLM models with a mix of general code-based, paid, and open-source models, namely ChatGPT 4, ChatGPT 4o, LLaMA 8B, LLaMA 13B, CodeLLaMA, and Mistral. First, we selected 100 scripts from the collected build scripts.  Next, we applied zero-shot prompting across six LLMs, instructing them to analyze the build scripts and identify instances of code smells. We provided an initial list of potential code smells identified in our qualitative analysis of GitHub issues.
The models were tasked with detecting both the predefined code smells and any additional, previously unrecognized code smells they could infer from the scripts. Prompt details are available in data availability~\cite{datashare}.

\textit{\textbf{Validation of LLM:}}
To evaluate the models' performance, three authors independently and manually analyzed a selection of 100 build scripts to verify detected code smells.
The results from this manual analysis were then compared against the results provided by the six LLMs' outputs. To avoid potential bias, the first author was responsible for executing the LLMs and comparing their output with the manual analysis, while the second, third, and fourth authors conducted the manual evaluations and resolved any conflict through iterative discussion and following negotiation agreement~\cite{campbell2013coding}.
Among the analyzed LLMs, ChatGPT-4o showed the highest recall, accurately identifying the majority of true positive code smells, presented in Table~\ref{LLM_Comp}. As a result, ChatGPT-4o was selected for further analysis.

\textit{\textbf{Code Smell Taxonomy:}}
We analyzed 2,134 collected build scripts using ChatGPT-4o, following a consistent prompt and procedure.
This process provided a large list of code smells.
To facilitate structured analysis, we grouped similar code smells into broader smell categories.
For instance, LLM-generated responses such as ``HTTP URL for Maven Central," ``Insecure Repository URL," and ``HTTP Repository URL" were grouped under the category ``Insecure URLs," as they reflect a common pattern of using unsecured links within build scripts. We merged findings from our GitHub repository analysis with the LLM-generated results, ensuring that the final taxonomy was grounded in both automated detection and manual findings. Two coders independently conducted the analysis, with iterative refinement until thematic saturation was reached~\cite{guest2006many}.
Through this process, we identified 13 distinct code smell categories. The inter-rater reliability for categorization was a Cohen’s \texttt{k} of 0.81, which indicates strong agreement between coders. The final and complete list of identified categories is provided in Table~\ref{CS-CWE}.


\begin{table}
\caption{LLMs Comparison Based on Recall}\label{LLM_Comp}
\resizebox{\columnwidth}{!}{%
\begin{tabular}{|l|l|l|l|l|}
\hline
           & Maven & Gradle & CMake & Make  \\ \hline
Llama 8B   & 41\% & 36\%  & 27\% & 13\% \\ \hline
Llama 13B  & 32\% & 38\%  & 20\% & 13\% \\ \hline
Codellama  & 29\% & 34\%  & 14\% & 16\% \\ \hline
Mistral    & 41\% & 43\%  & 31\% & 13\% \\ \hline
ChatGPT 4  & 59\% & 52\%  & 36\% & 33\% \\ \hline
ChatGPT 4o & 78\% & 75\%  & 52\% & 42\% \\ \hline
\end{tabular}%
}
\end{table}

\subsubsection{\textbf{Code Smells to CWE Mapping}} 
To assess the security relevance of the identified code smells in build scripts, we mapped each instance to corresponding entries in the Common Weakness Enumeration (CWE) database~\cite{CWE}. We selected CWE as the reference framework due to its standardized taxonomy of software security weaknesses and proper maintenance by the cybersecurity community. For example, the smell ``Hardcoded credential'' was mapped to both CWE-798: Use of Hard-coded Credentials and CWE-259: Use of Hardcoded Password~\cite{CWE}. To be consistent with the objectivity in the mapping process, two authors independently performed mapping between code smells and CWEs. Further, we resolved our conflicts through discussion with another author. Our evaluations demonstrate perfect consistency, resulting in a Cohen’s Kappa score of 1.0 after resolving conflict, which indicates perfect inter-rater reliability. A list of the identified code smells across different types of build scripts with the associated CWE is in Table~\ref{CS-CWE}.

\begin{table*}

\caption{List of Identified Code Smell}\label{CS-CWE}
\resizebox{\textwidth}{!}{
\begin{tabular}{|l|l|l|l|l|l|}
\hline
\multicolumn{1}{|c|}{\textbf{Code Smell Category}}                         & \multicolumn{1}{c|}{\textbf{Maven}} & \multicolumn{1}{c|}{\textbf{Gradle}} & \multicolumn{1}{c|}{\textbf{CMake}} & \multicolumn{1}{c|}{\textbf{Make}} & \multicolumn{1}{c|}{\textbf{Common Weakness Enumeration (CWE)}}                                                                                    \\ \hline
Complexity                                                                 &                                     &                                      & *                                   & *                                  & CWE-710: Improper Adherence to Coding Standards                                                                                                    \\ \hline
Deprecated Dependencies                                                    & *                                   & *                                    & *                                   & *                                  & CWE-1104: Use of Unmaintained Third-Party Components                                                                                               \\ \hline
Duplicate                                                                   & *                                   & *                                    & *                                   & *                                  & CWE-710: Improper Adherence to Coding Standards                                                                                                    \\ \hline
Empty/Incomplete Tags                                                       & *                                   & *                                    &                                     &                                    & CWE-611: Improper Restriction of XML External Entity Reference (XXE)                                                                               \\ \hline
Hardcoded Credentials                                                      & *                                   & *                                    & *                                   & *                                  & \begin{tabular}[c]{@{}l@{}}CWE-798: Use of Hard-coded Credentials\\ CWE-259: Use of Hard-coded Password\end{tabular}                               \\ \hline
Hardcoded Paths/ URLs                                                       & *                                   & *                                    & *                                   & *                                  & \begin{tabular}[c]{@{}l@{}}CWE-427: Uncontrolled Search Path Element\\ CWE-706: Use of Incorrectly-Resolved Name or Reference\end{tabular}         \\ \hline
Inconsistent Dependency Management                                          & *                                   & *                                    &                                     &                                    & \begin{tabular}[c]{@{}l@{}}CWE-439: Behavioral Change in New Version or Environment\\ CWE-710: Improper Adherence to Coding Standards\end{tabular} \\ \hline
Insecure URLs                                                              & *                                   & *                                    & *                                   & *                                  & CWE-319: Cleartext Transmission of Sensitive Information                                                                                           \\ \hline
Lack Error Handling                                                         & *                                   & *                                    & *                                   & *                                  & CWE-391: Unchecked Error Condition                                                                                                                  \\ \hline
Missing Dependency Version                                                  & *                                   & *                                    & *                                   & *                                  & CWE-440: Expected Behavior Violation                                                                                                                \\ \hline
Outdated Dependencies                                                      & *                                   & *                                    & *                                   & *                                  & CWE-1104: Use of Unmaintained Third-Party Components                                                                                               \\ \hline
Suspicious Comments                                                         & *                                   & *                                    & *                                   & *                                  & CWE-546: Suspicious Comment                                                                                                                        \\ \hline
Wildcard Usage                                                              & *                                   & *                                    & *                                   & *                                  & CWE-829: Inclusion of Functionality from Untrusted Control Sphere                                                                                   \\ \hline
\end{tabular}
}
\end{table*}

\subsection{Answer to RQ1: What code smells occur in build scripts?}~\label{RQ1-res}
In this section, we provided an answer for RQ1. While some of these categories overlap with smells identified in prior studies, Infrastructure-as-Code (IaC)~\cite{rahman2019seven} and CI/CD pipelines~\cite{delicheh2024mitigating,pan2023ambush,zhang2022buildsonic}, our findings reveal how these issues manifest uniquely in build scripts. Below, we discussed the identified thirteen code smell categories in detail. An annotated Maven script illustrating all categories is shown in Figure~\ref{maven_example}.
\begin{itemize}
\item \textbf{Complexity:} This code smell refers to overly complex logic in build scripts, such as nested conditionals, inline shell commands, or convoluted plugin chains. This smell is also known as conditional complexity, using lengthy, cascading if statements or switch/case~\cite{complexity}. The existence of such smells reduces maintainability and increases the likelihood of misconfiguration. This smell is linked to CWE-710: Improper Adherence to Coding Standards.
\item \textbf{Deprecated Dependencies:} This smell refers to the recurring pattern of using abandoned or not-maintained dependencies. Using deprecated libraries or APIs in build configurations is indicative of poor dependency management~\cite{miller2025understanding}. Deprecated components are typically no longer maintained or updated and may harbor known vulnerabilities. This smell is closely associated with CWE-1104: Use of Unmaintained Third-Party Components, which refers to the risks of depending on obsolete software packages. 
\item \textbf{Duplicates:} Duplicate code declarations occur when the same dependency, code, or configuration is redundantly included in the build script~\cite{li2019dlfinder, Duplicate}. This practice leads to bloated scripts, increases maintenance overhead, and may cause unexpected behaviors due to overriding rules. Although there is no exact CWE that captures this issue, it falls under the CWE-710: Improper Adherence to Code Standards, which represents poor coding practices.
\item \textbf{Empty/Incomplete Tags:} This smell refers to the use of XML elements without content (e.g., \texttt{<modelVersion></modelVersion>}), which can lead to undefined behavior or misinterpretation by tools. This smell is linked to  CWE-611: Improper Restriction of XML External Entity Reference (XXE), indicating a lack of standardization. As this is an XML-based code smell, among the four build systems we studied, this code smell appeared only in Maven and Gradle.
\item \textbf{Hardcoded Credentials:} Hardcoded credentials refer to the recurring pattern of embedding authentication or sensitive information, such as usernames, passwords, API tokens, or private keys, directly within build scripts. This approach poses a major security risk as sensitive information may be inadvertently exposed through version control. As a result, secrets can be exposed to public repositories and cause unauthorized access to the system. Prior work shows this smell in IaC and CI/CD (e.g., secrets in Puppet, workflow tokens), with prolonged lifetimes and high exploitation risk~\cite{rahman2019seven,pan2023ambush}. In build scripts, the risk is amplified because credentials are embedded in artifact management and dependency resolution workflows, directly affecting supply-chain integrity. This smell aligns with CWE-798: Use of Hard-coded Credentials and CWE-259: Use of Hard-coded Password, which highlights the vulnerability of storing sensitive data in an unprotected and immutable manner.
\item \textbf{Hardcoded Paths/URLs:} This smell refers to the recurring pattern of the direct inclusion of absolute paths or fixed URLs in build scripts. Embedding absolute paths or fixed URLs directly into scripts can lead to failures when the code is executed in different environments, as these paths may not exist or may differ across systems. This could reduce the portability and adaptability of the build process~\cite{hardcodedpath,rahman2019source}.
Furthermore, hardcoded network paths can expose the system to untrusted locations. This smell aligns with CWE-427: Uncontrolled Search Path Element and CWE-706: Use of Incorrectly-Resolved Name or Reference.
\item \textbf{Inconsistent Dependency Management:} This smell refers to the inconsistency, such as simultaneous use of hardcoded versions and version variables within the same project or using different versions of the same dependency within the same code block. Such inconsistency can result in dependency conflicts and undermine reproducibility. While not directly to a specific CWE, this issue aligns with broader software quality concerns, such as CWE-440 Expected Behavior Violation.

\item \textbf{Insecure URLs:} Using non-secure URLs (i.e., HTTP instead of HTTPS) for fetching dependencies or uploading artifacts is considered a code smell. This practice exposes communication channels to man-in-the-middle (MITM) attacks and tampering~\cite{rescorla2000rfc2818}. Similar misconfigurations in CI/CD have been studied~\cite{delicheh2024mitigating,pan2023ambush}, but our analysis shows that insecure URLs are especially widespread in Maven (93\% of scripts). This finding suggests that insecure URLs are not isolated errors but a systemic, tool-driven pattern in build scripts, contributing to long-term security debt in software supply chains. This smell corresponds to CWE-319: Cleartext Transmission of Sensitive Information.

\item \textbf{Lack of Error Handling:} 
For build scripts, this smell refers to a lack of error checking or passing the code with an error. For example, if the maven script ``sql-maven-plugin'' is configured with \texttt{<onError>continue</onError>}, which means that the build will continue even if errors occur during SQL execution. This can mask issues that should be addressed before deployment. It is a code smell because it can lead to overlooking major errors, resulting in unstable or incorrect build artifacts. This smell falls under CWE-391: Unchecked Error Condition. 

\item \textbf{Missing Dependency Version:} When a dependency is declared without an explicit version number, the build system applies different dependency resolution mechanisms. For example, Maven will fail the build unless a version is inherited~\cite{mavenfail} and Gradle attempts to resolve the dependency via transitive dependencies~\cite{gradlefail}, which potentially introduces unstable or insecure components. This form of version drift is a common software supply chain risk and is covered under CWE-439: Behavioral Change in New Version or Environment, and
CWE-710: Improper Adherence to Coding Standards.

\item \textbf{Outdated Dependencies:} This smell refers to a practice when build scripts reference older versions of dependencies for which more secure or stable releases exist. Persisting with outdated libraries increases the likelihood of exposure to known vulnerabilities and creates an attack vector~\cite{lauinger2018thou}. As with deprecated components, this issue is categorized under CWE-1104, emphasizing the need for timely dependency upgrades.

\item \textbf{Suspicious Comments:} This smell refers to comments that expose unresolved defects, missing functionality, or potential weaknesses in the system. Similar to prior work on IaC~\cite{rahman2019seven}, we also observed this smell in build scripts. It corresponds to CWE-546: Suspicious Comment. Typical examples include annotations such as “TODO” or “FIXME,” which often signal latent issues that have not yet been addressed~\cite{storey2008todo}. 

\item \textbf{Wildcard Usage:} This code smell refers to not specifying the version numbers for the build, instead using `*' or `+'. The given example in Figure~\ref{maven_example} shows that ``software.amazon.awssdk:*'' has been used. During prototyping or early development, using the wildcard version helps in the fast inclusion of all artifacts from certain groups, but in the long run, this can cause version drift~\cite{versiondrift}, dependency bloat~\cite{soto2021longitudinal}, or dependency confusion~\cite{wang2018dependency}. This smell is mapped to CWE-829: Inclusion of Functionality from Untrusted Control Sphere.

\end{itemize}
\begin{figure}
    \centering
\fbox{\includegraphics[width=\linewidth]{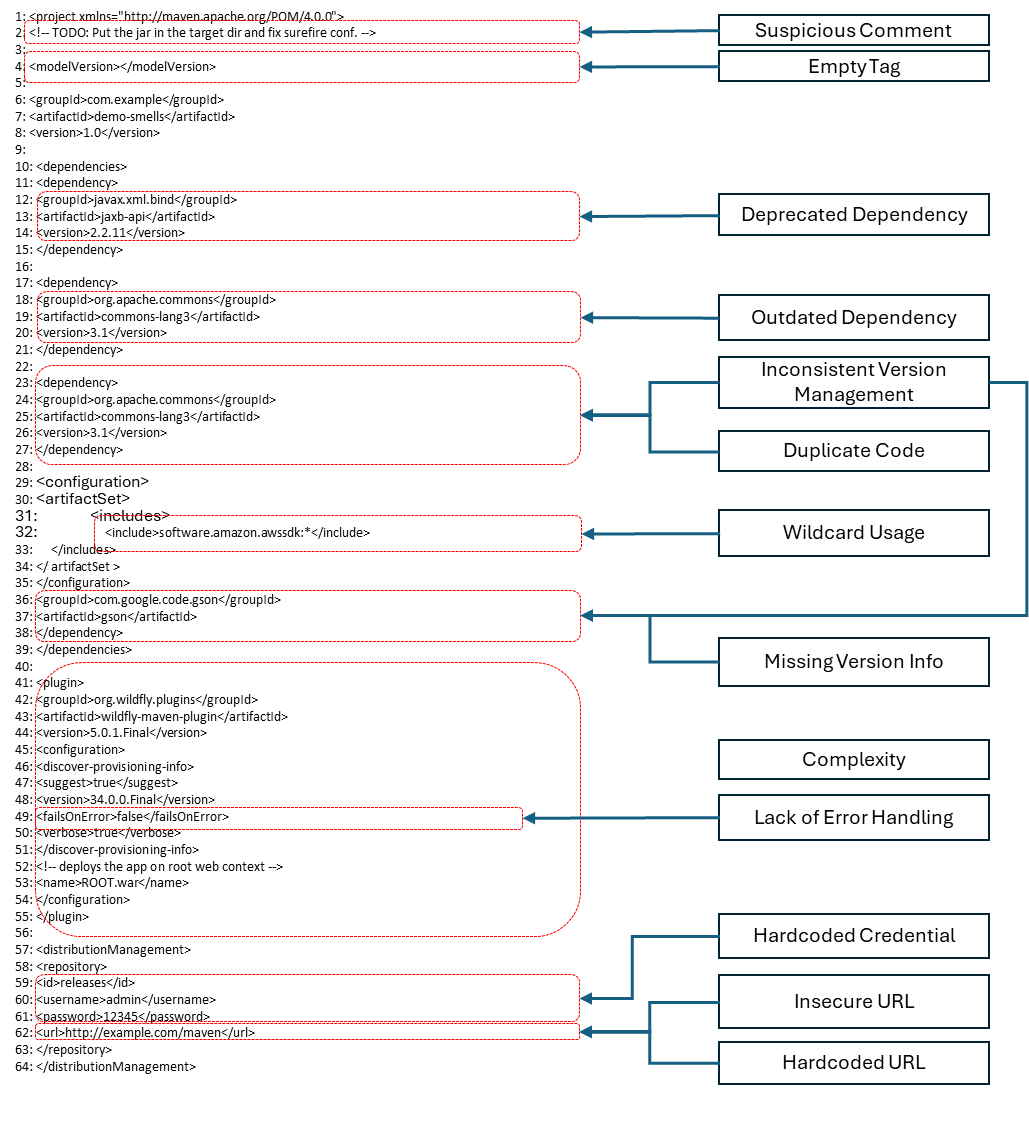}}
    \caption{An annotated Maven script example with all identified code smell categories.}~\label{maven_example}
    \label{maven}
\end{figure}

\section{Sniffer: Static Analytics Tool for Build Scripts}\label{sniffer_development}
Our static analysis tool, \textit{Sniffer}, has been developed to automatically detect code smells in build scripts. In this section, we describe the development and evaluation process of Sniffer's smell detection. Our tool is available in data availability~\cite{datashare}.

\subsection{\textbf{Design and Development of Sniffer}}
\textit{Sniffer} is a static‑analysis tool that detects security and
maintainability-oriented code smells in four widely‑used build systems:
Maven, Gradle, CMake, and Make. The following paragraphs describe each layer in turn.

\subsubsection*{1.~Input Orchestration}
On invocation, the dispatcher determines the target build system by (i)
examining the file name and (ii) applying lightweight lexical
heuristics, for example, searching the first kilobyte for
\textit{cmake\_minimum\_required} or the presence of make‑style rules.  The file is then forwarded to the corresponding parser module.

\subsubsection*{2.~Specialised Parsers}
\begin{description}[leftmargin=1.4em,labelsep=0.4em]
  \item[Maven.]  POM files are loaded as raw bytes, any editor‑added line
        prefixes are stripped, and the payload is parsed with
        \textit{lxml.XMLParser(recover=true)}.  The resulting tree is
        normalized into lists of dependencies, plugins, and XML nodes used
        by downstream checks.
  \item[Gradle.]  A tri‑regex strategy extracts dependency declarations in
        Groovy DSL, Kotlin DSL, and map notation; repository URLs are
        captured via a dedicated pattern.  The full script is preserved in
        \texttt{raw\_content} for textual analyses.
  \item[CMake.]  The parser identifies \textit{find\_package},
        \textit{add\_executable}, and \textit{add\_library} directives and
        records their arguments.  Files lacking an explicit
        \textit{cmake\_minimum\_required} statements are still analyzed,
        improving coverage of legacy projects.
  \item[Make.]  To avoid side effects, files containing
        \verb|$(shell ...)| fall back to a naïve line scanner that
        collects variable assignments, rules, and commands.  Otherwise,
        the parser exploits \texttt{make -pn} to obtain an expanded
        database, which is then converted into a structured
        \texttt{MakefileAnalysis} object.
\end{description}

\subsubsection*{3.~Shared Utility Layer}
Common functionality is consolidated in \texttt{common/}.  In particular,
\texttt{version\_utils.py} performs metadata look‑ups on Maven Central and
implements semantic‑version comparison.  Results are cached in an
LRU store to bound network overhead.

\subsubsection*{4.~Dynamic Rule Engine}
All smell detectors are defined as functions whose names begin with
\verb|check_| and that accept exactly one positional argument.  At start‑up
the engine introspects every \textit{security‑check} module, registers the
matching callables, and executes them in sequence.  Each detector returns a
set of issue dictionaries of the form
\{\texttt{smell\_id},~\texttt{issue},~\texttt{severity}\}; exceptions are
captured and re‑emitted as low‑severity findings, ensuring that the individual
failures do not interrupt analysis.

\subsubsection*{Smell Catalog}
Forty‑two detectors are implemented to detect the smells.  They span version‑related
risks (missing, inconsistent, outdated, or stale dependencies), hard‑coded
secrets and absolute paths, insecure transport (HTTP URLs), duplicate or
unused declarations, complexity heuristics, and inadequate error handling.
All smells are mapped to relevant CWEs to facilitate security triage.  


\subsubsection*{Extensibility}
Supporting an additional build system entails providing (i) a parser that
emits the canonical data schema and (ii) optional smell detector modules.
The dispatcher and rule engine remain untouched, offering a clear path for
future expansion (e.g.,\ Bazel or SCons). Details of each of the four build types: parser pattern, regular expressions, the ruleset, and detection logic used behind Sniffer are provided~\cite{datashare}.

\medskip
\noindent
\subsection{\textbf{Performance Evaluation}}\label{lint_evaluation}
In this section, we discussed the steps of Sniffer's performance evaluation. We evaluated our tool's performance in two ways. i) Evaluation Against Manual Analysis ii) Evaluation Against Oracle Dataset. Both approaches are explained below. The results of both analyses are discussed in Section~\ref{evaluation_res}.
\subsubsection{Evaluation Against Manual Analysis}~\label{evaluation_manual}

To test Sniffer’s performance, we first randomly selected 100 scripts collected from four datasets. Next, two authors manually analyzed the scripts for each of the identified code smells. We then ran Sniffer on the same sampled build scripts and measured the Precision, Recall, and F1 values. Next, we compared Sniffer’s output with our manual analysis to evaluate the tool’s performance.
Our findings are presented in Table~\ref{performance}.
\subsubsection{Evaluation Against Oracle Dataset}~\label{evaluation_oracle}
We constructed the Oracle dataset using a closed coding approach~\cite{closedcode}, where a set of raters analyzes material and identifies code smells based on a predefined codebook~\cite{datashare}. The material consisted of 72 scripts that were manually examined for code smells. Raters applied their knowledge of programming and build scripts to determine whether a particular smell was present in each script. To avoid bias, raters were not involved in this research as part of the primary code smells identification or the development of Sniffer. 
As raters, we recruited 20 Computer Science graduate students from NC State University who had some level of familiarity with build scripts to serve as raters. We obtained an institutional review
board (IRB) approval for student participation~\ref{ethics}. Each rater was compensated with a \$20 gift card for their involvement. The 72 scripts were sampled to include a mix of scripts with and without code smells.
These scripts were then distributed among the 20 raters, ensuring that each script was independently reviewed by at least two raters, with no rater reviewing more than eight scripts.
The smell identification task was submitted through Qualtrics. In each task, a rater determined which of the code smells identified in Section III were present in a given script.
We observe agreements on the
scripts, with a Cohen’s Kappa of 0.63. According to
Artstein and Poesio's interpretation~\cite{artstein2008inter}, the reported agreement
is considered ``substantial''. After the construction of the Oracle dataset, we evaluated Sniffer's performance by comparing Oracle results with Sniffer's findings using Precision, Recall, and F1 scores. 
\\
\subsection{Performance Evaluation Results}~\label{evaluation_res}
Sniffer's performance for Precision, Recall, and F1 scores is presented in Table~\ref{performance}. As shown in the table, the Precision, Recall, and F1 values of the developed tool were measured against the manual analysis, which served as the ground truth. Additionally, we evaluated the tool against the oracle's output by measuring precision and recall across four types of build scripts: Maven, Gradle, CMake, and Make.
When compared against manual analysis, Sniffer achieved consistently high performance, with an average precision of 0.89, a recall of 0.88, and an F1-score of 0.883. Among individual build systems, Maven achieved the highest precision (0.92), while CMake showed the strongest recall (0.93) and F1-score (0.920), indicating reasonably high agreement with ground truth. We also measured the confidence intervals. The 95\% confidence interval for manual analysis is (0.852–0.914), indicating a narrow range around the mean, reflecting stable performance. In terms of performance evaluation against the Oracle, it showed slightly lower but comparable results, with an average precision of 0.83, a recall of 0.81, and an F1-score of 0.819. Maven again achieved the highest precision (0.84), while both Gradle and CMake obtained the top recall values (0.81). The 95\% confidence interval for Oracle evaluation (0.807–-0.829) further confirms the consistency of these results. Overall, Sniffer demonstrated strong agreement with manual analysis and reliable performance against the Oracle dataset. The narrow confidence intervals for both comparisons highlight the stability of the tool’s results, underscoring Sniffer’s potential as an effective and reliable automated approach for detecting code smells in build scripts.

\begin{table}
\caption{Sniffer's Precision, Recall, and F1 Scores}\label{performance}
\resizebox{\columnwidth}{!}{%
\begin{tabular}{|l|lll|lll|}
\hline
\multicolumn{1}{|c|}{\textbf{}} & \multicolumn{3}{c|}{\textbf{Manual Analysis}} & \multicolumn{3}{c|}{\textbf{Oracle}} \\ \hline
\textbf{Build Scripts} & \textbf{Precision} & \textbf{Recall} & \textbf{F1} & \textbf{Precision} & \textbf{Recall} & \textbf{F1} \\ \hline
Maven  & 0.92 & 0.83 & 0.874 & 0.84 & 0.83 & 0.835 \\
Gradle & 0.90 & 0.89 & 0.895 & 0.81 & 0.81 & 0.810 \\
CMake  & 0.91 & 0.93 & 0.920 & 0.82 & 0.81 & 0.815 \\
Make   & 0.83 & 0.86 & 0.845 & 0.83 & 0.80 & 0.815 \\ \hline
\textbf{Average} & 0.89 & 0.88 & 0.883 & 0.83 & 0.81 & 0.819 \\ \hline
\end{tabular}
}
\end{table}

\section{Empirical Analysis of Build Scripts}\label{em_analysis}
In this section, we provided an answer for RQ2. We performed an empirical analysis of code smell in build scripts using \textit{Sniffer}. We discuss the build script collection method in Section~\ref{build-data} and the findings for RQ2 in Section~\ref{em-result}.
\subsection{RQ2: How frequently do code smells occur in build scripts?}\label{build-data}

\subsubsection{Dataset}
In this phase of the study, we conducted an empirical study with a large-scale dataset of Maven, Gradle, CMake, and Make scripts. To examine the prevalence of identified code smells and enhance the generalizability of our results, we focused on GitHub repositories, which are widely used by organizations to host prominent open-source software (OSS) projects~\cite{vidoni2022systematic}.
In alignment with established research practices~\cite{munaiah2017curating}, we focused on collecting OSS repositories to have a diverse and accessible dataset. The  data collection process was followed by a set of pre-defined inclusion criteria, detailed as follows:
\begin{itemize}
    \item Criteria 1: The repository should contain at least one of the selected types: Maven, Gradle, CMake, Make. The Repositories should contain any of the following types: pom.xml, build.gradle, build.gradle.kts, CMake.txt, or Makefile.
    \item Criteria 2: The repositories are not clones or duplicates.
 
\end{itemize}
To avoid redundancy, repositories that have already been processed and recorded in a central Google Sheet in Section~\ref{Build_QA} data collection were skipped.
This automated approach allowed us to collect a large and diverse set of build scripts from GitHub, forming the basis for the evaluation of the static analysis tool and our empirical study.
We answer RQ2 using 5882 scripts
collected from 913 (Maven), 634 (Gradle), 443 (CMake), and 2887 (Make) repositories, respectively. Summary attributes of the collected repositories are listed in Table~\ref{repo}. Since most repositories contained only a single build script, the total number of build scripts collected is approximately equal to the number of repositories for Maven, Gradle, and Makefiles. Moreover, given the longstanding use of Makefiles in software development, the majority of the collected data consisted of Makefiles.
\begin{table}
\caption{Summery of Collected Repositories}\label{repo}
\resizebox{\columnwidth}{!}{%
\begin{tabular}{|l|llll|}
\hline
\textbf{Attributes} & \multicolumn{4}{c|}{\textbf{Values}}                                                         \\ \hline
Script Type         & \multicolumn{1}{c|}{Maven} & \multicolumn{1}{l|}{Gradle} & \multicolumn{1}{l|}{CMake} & Make \\ \hline
Repositories        & \multicolumn{1}{l|}{913}   & \multicolumn{1}{l|}{634}    & \multicolumn{1}{l|}{443}   & 2887  \\ \hline
Files               & \multicolumn{1}{l|}{918}      & \multicolumn{1}{l|}{634}       & \multicolumn{1}{l|}{443}      &3887      \\ \hline
Total LOC           & \multicolumn{1}{l|}{125502}      & \multicolumn{1}{l|}{30872}       & \multicolumn{1}{l|}{23679}      & 636283     \\ \hline
\end{tabular}
}
\end{table}

\subsection{Answer to RQ2: How frequently do code smells occur in build scripts?}\label{em-result}
\subsubsection{Occurrence}
The occurrence is the number of each code smell across different build systems. As represented in Table~\ref{occr}, among all smells, Wildcard Usage was the most frequent, with 2,205 instances in Make scripts. Insecure URLs are followed, especially common in Maven (854) and Make (587). Lack of Error Handling also stood out in Make (988), while Inconsistent Dependency Management was largely observed in Maven (797). Additionally, Hardcoded Paths/URLs appeared frequently in Make (664) and Maven (373), and Suspicious Comments were notable in Make (685) and Gradle (104). Deprecated Dependencies were dominant in Gradle (217), and Outdated Dependencies in Maven (319). Duplicate entries were mostly found in Maven (139) and Gradle (116), with none in Make.

Other smells like Hardcoded Credentials, Missing Dependency Version, and Complexity also showed varying occurrences, with Make and Gradle scripts often reporting higher counts.
The ``no smells'' represents the number of scripts where no code smell was found for that certain build type. These patterns suggest that certain smells are more prevalent in specific ecosystems, possibly due to differing development practices or code structure.
\subsubsection{Proportion of script}
\begin{itemize}
\item Approach:  The proportion of scripts metric indicates the prevalence of a smell across individual scripts~\cite{rahman2019seven}. This metric reflects the percentage of scripts that contain at least one occurrence of smell.

\item Results: As presented in Table~\ref{occr}, Insecure URLs appeared in the highest proportion of scripts, up to 93\% in Maven and 15.1\% in Make, underscoring their widespread presence and associated security risks. Wildcard Usage was also highly prevalent, particularly in Make (56.7\%) and CMake (23.3\%), suggesting potentially ambiguous dependency declarations. In contrast, Hardcoded Credentials were rare, occurring in only 0.9\% of Maven scripts and in less than 0.1\% across other types. 
\end{itemize}

\subsubsection{Smell Density}
\begin{itemize}
    \item Approach: In previous studies, researchers utilized vulnerability density~\cite{alhazmi2005quantitative} and defect density~\cite{rahmani2010study,nagappan2005use} to measure the prevalence of such problems. In line with the same concept, we employed equation~\ref{eq:1} to quantify the density of a code smell ($d$). Here, $x$ is the total occurrence of code smell for every 1000 Lines Of Code (LOC), represented by KLOC.
\begin{align}
    \text{Smell Density(d)}=\frac{\text{Total occurrences of x}}{\text{KLOC}}\label{eq:1}\\ 
    \text{where KLOC = LOC/1000} \nonumber
\end{align}

    \item Results: The Smell Density (per KLOC) column in Table~\ref{occr}  reports the frequency of security smells normalized per thousand lines of code (KLOC). Smell density provides a relative measure of smell intensity. Gradle scripts exhibited notably higher smell densities across several categories, such as Deprecated Dependencies (7.029 per KLOC), Duplicate (3.757), and Suspicious Comments (3.369), suggesting these scripts are often smaller but more smell-prone per unit of code. Conversely, Make scripts, while having high absolute occurrences, showed lower density for most smells (e.g., Outdated Dependencies: 0.030 per KLOC), indicating their higher LOC base dilutes the relative impact of smells. This highlights the importance of using density alongside raw counts for fair cross-tool comparisons.

\end{itemize}
\subsubsection{Smell co-occurrence matrix:}
\begin{itemize}
    \item Approach: To investigate how often one code smell is present with another code smell in build scripts, we performed a pairwise co-occurrence analysis. We calculated the percentage of each smell type $cs_i$ with another smell type $cs_j$ across our dataset of build scripts. Following the methodology proposed by the previous studies~\cite{yamashita2013exploring,palomba2018large}, we measured the co-occurrence by using the following formula:
\begin{equation*}
\text{Co-occurrence}_{cs_i \rightarrow cs_j} = \frac{|cs_i \cap cs_j|}{|cs_i|}  {where, i \neq j}
\end{equation*}
The equation,  ${|cs_i \cap cs_j|}$ represents the number of build scripts that contain both smell types $cs_i$ and  $cs_j$, and $|cs_i|$ denotes the total number of scripts containing $cs_i$. This directional metric captures the likelihood that the presence of smell $cs_i$ implies the presence of smell $cs_j$, enabling the identification of strongly associated smell pairs. Notably, $\text{Co-occurrence}_{cs_i \rightarrow cs_j} \neq \text{Co-occurrence}_{cs_j \rightarrow cs_i}$ due to the asymmetry of the denominator. 
    \item Results: As shown in the Figure~\ref{heatmap}, the code smell co-occurrence in build scripts has several strong associations between specific pairs of smells. Empty/Incomplete Tags exhibit perfect co-occurrence with Insecure URLs (1.00) and a near-perfect association with Inconsistent Dependency Management (0.98), indicating that structurally deficient scripts often suffer from insecure configurations and poor dependency practices. Similarly, Duplicates co-occur frequently with both Hardcoded Paths/URLs (0.67) and Insecure URLs (0.68), suggesting that redundancy in script elements is commonly linked with insecure or non-modular path specifications.
\end{itemize}

\begin{table*}
\caption{Summarization of Smell Occurrences, Smell Density, and Proportion of Scripts for the Four Build Types}\label{occr}
\resizebox{\textwidth}{!}{%
\begin{tabular}{l|llll|llll|llll}
\hline
\multicolumn{1}{c|}{\textbf{Code Smell Name}} & \multicolumn{4}{c|}{\textbf{Occurrence}} & \multicolumn{4}{c|}{\textbf{Smell Density (per KLOC)}} & \multicolumn{4}{c}{\textbf{Proportion of Script}} \\ \hline
                                              & Maven & Gradle & CMake & Make & Maven & Gradle & CMake & Make & Maven & Gradle & CMake & Make \\ \hline
Complexity                                    & 11    & 99     & 37    & 504  & 0.088 & 3.207  & 1.563 & 0.792 & 0.012 & 0.156  & 0.084 & 0.130 \\ \hline
Deprecated Dependencies                       & 95    & 217    & 11    & 13   & 0.757 & 7.029  & 0.465 & 0.020 & 0.103 & 0.342  & 0.025 & 0.003 \\ \hline
Duplicates              & 139   & 116    & 19    & 0    & 1.108 & 3.757  & 0.802 & 0.000 & 0.151 & 0.183  & 0.043 & 0.000 \\ \hline
Empty/Incomplete Tags            & 41    & 0      & 0     & 0    & 0.327 & 0.000  & 0.000 & 0.000 & 0.045 & 0.000  & 0.000 & 0.000 \\ \hline
Hardcoded Credentials                         & 8     & 47     & 3     & 217  & 0.064 & 1.522  & 0.127 & 0.341 & 0.009 & 0.074  & 0.007 & 0.056 \\ \hline
Hardcoded Paths/URLs                       & 373   & 279    & 25    & 664  & 2.972 & 9.037  & 1.056 & 1.044 & 0.406 & 0.440  & 0.056 & 0.171 \\ \hline
Inconsistent Dependency Management            & 797   & 19     & 0     & 0    & 6.350 & 0.615  & 0.000 & 0.000 & 0.868 & 0.030  & 0.000 & 0.000 \\ \hline
Insecure URLs                                 & 854   & 123    & 27    & 587  & 6.805 & 3.984  & 1.140 & 0.923 & 0.930 & 0.194  & 0.061 & 0.151 \\ \hline
Lack Error Handling            & 303   & 4      & 25    & 988  & 2.414 & 0.130  & 1.056 & 1.553 & 0.330 & 0.006  & 0.056 & 0.254 \\ \hline
Missing Dependency Version                    & 121   & 24     & 41    & 278  & 0.964 & 0.777  & 1.731 & 0.437 & 0.132 & 0.038  & 0.093 & 0.072 \\ \hline
Outdated Dependencies                         & 319   & 87     & 18    & 19   & 2.542 & 2.818  & 0.760 & 0.030 & 0.347 & 0.137  & 0.041 & 0.005 \\ \hline
Suspicious Comments            & 114   & 104    & 28    & 685  & 0.908 & 3.369  & 1.182 & 1.077 & 0.124 & 0.164  & 0.063 & 0.176 \\ \hline
Wildcard Usage                                & 9     & 142    & 103   & 2205 & 0.072 & 4.600  & 4.350 & 3.465 & 0.010 & 0.224  & 0.233 & 0.567 \\ \hline
\textbf{No Smells}                            & 34    & 177    & 251   & 1151 &       &        &       &       &       &        &       &       \\ \hline
\end{tabular}
}
\end{table*}

\begin{figure*}[!t]
    \centering
    \includegraphics[width=.70\linewidth]{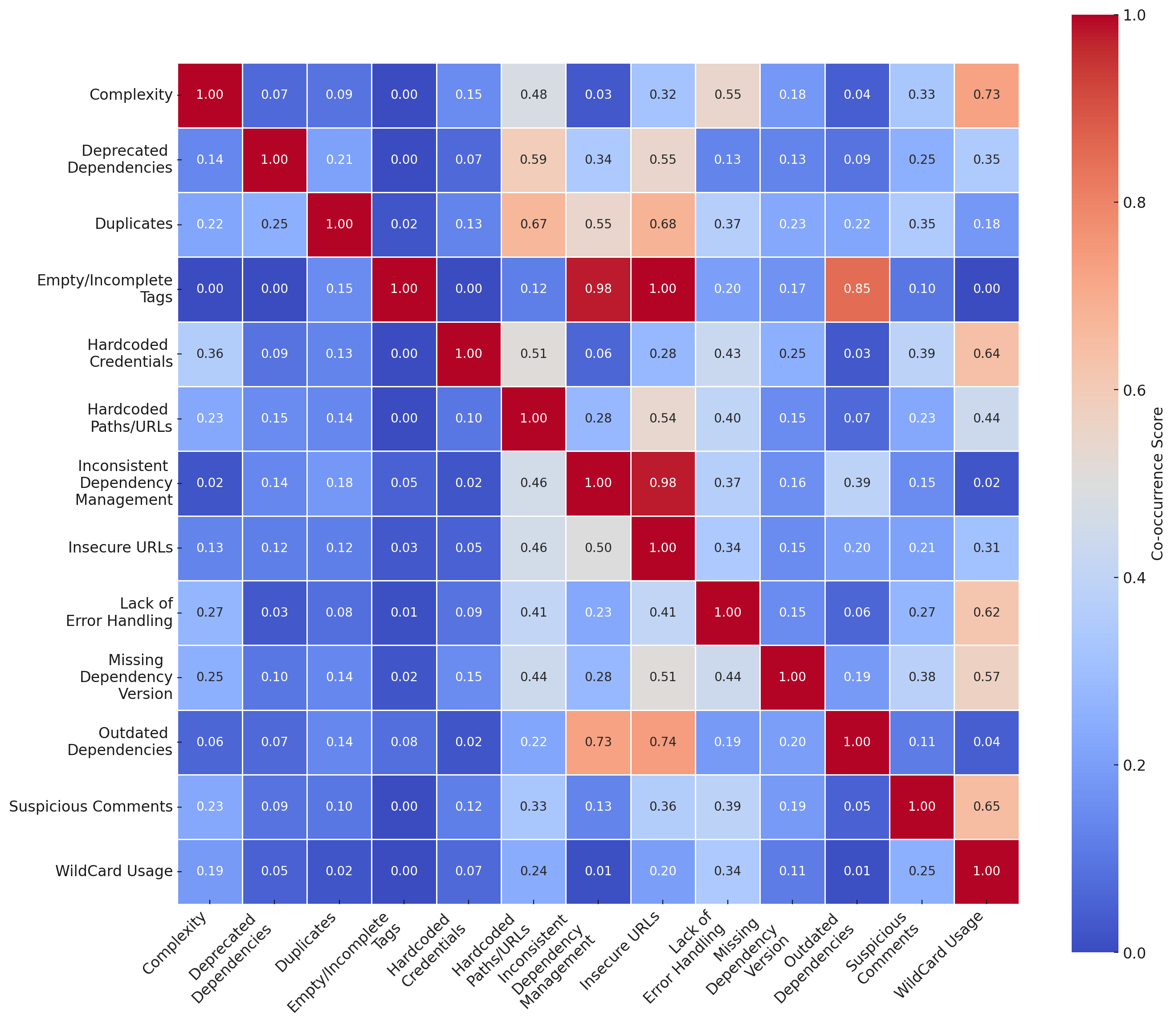}
    \caption{Heatmap Of Code Smell Co-Occurrence In Build Scripts}
    \label{heatmap}
\end{figure*}
\section{Discussion}\label{discussion}
In this section, we discuss and recommend mitigation strategies for the identified security smells in the build script in Section~\ref{mitigation}, discuss the implementation of Sniffer for secure build script practice in ~\ref{implication}, and provide guidelines for future work in Section~\ref{future-work}. 

\subsection{Mitigation Strategies}~\label{mitigation}
The identification of recurring code smells within build scripts highlights key areas where software quality, maintainability, and security can be significantly improved. As tools like Sniffer can help in code smell detection, mitigation is also essential for a secure build system. In this section, we discuss and recommend strategies to mitigate each of the identified code smells.

\textbf{Complexity:}
To address complexity in scripts, development teams should adopt clear coding conventions and modularize build logic into reusable components or scripts. Regular refactoring sessions and peer reviews help manage complexity, making build scripts easier to audit and maintain.

\textbf{Deprecated Dependencies:}
Regular dependency monitoring and automated tools such as OWASP Dependency-Check or utilizing Software Composition Analysis (SCA) should be employed to identify deprecated or unmaintained libraries. The software development team should proactively replace deprecated dependencies with supported alternatives to minimize security risks~\cite{pashchenko2018vulnerable}.

\textbf{Duplicates:}
Mitigating duplicate declarations involves regular dependency audits and leveraging build tools (e.g., Maven’s dependency analysis plugins) to detect redundancy. Clean, simplified dependency structures enhance maintainability and reduce complexity.

\textbf{Empty/Incomplete Tags}
To mitigate empty tags or elements, automated validation tools (e.g., XML schema validators) should be integrated into CI/CD pipelines. Adherence to XML schema standards prevents unintended behavior and enhances script clarity.

\textbf{Hardcoded Credentials}
To mitigate this issue, developers should adopt secure credential management practices, such as utilizing dedicated secret management systems (e.g., AWS Secrets Manager or environment variables). Secrets must never be directly stored in scripts or repositories. Scanning repositories for exposed secrets using tools like TruffleHog, Gitleaks, or GitHub Secrets can help in secret leakage mitigation. In addition, credential scanning tools, such as \textit{Sniffer}, can automate the detection and remediation of hardcoded credentials.

\textbf{Hardcoded Paths and URLs}
Mitigation involves replacing absolute paths and URLs with relative paths, environment variables, or centralized configuration files. Such practices enhance portability and adaptability across diverse development environments and reduce security risks~\cite{martin2009clean}.

\textbf{Inconsistent Dependency Management:}
Adopting consistent version management through standardized properties, dependency management tools (e.g., Maven’s BOM feature), and unified policies significantly reduces inconsistency and simplifies dependency updates ~\cite{humble2010continuous}.

\textbf{Insecure URLs:}
Mitigating this involves enforcing HTTPS protocols for all external connections within build scripts. Automated scanning tools can continuously check build scripts for insecure HTTP links, thus preventing man-in-the-middle vulnerabilities.

\textbf{Lack of Error Handling:}
Lack of Error Handling is concerning in automated build environments where early detection and halting failure are essential to maintaining software quality. Continuing execution after an error may lead to partially configured systems or incomplete dependency states, increasing the risk of introducing latent vulnerabilities. To address this, concrete error-handling policies within build scripts are needed. A recommended mitigation strategy involves configuring plugins to fail explicitly upon encountering errors (e.g.,\texttt{<onError>fail</onError>} and \texttt{<failOnError>true</failOnError>}). In addition, enforcing a fail-fast mechanism can help~\cite{fastfail}. This approach not only aligns with recommended practices in continuous integration and DevSecOps but also supports reproducible and verifiable builds~\cite{sgibson}.

\textbf{Missing Dependency Version:}
Version specifications need to be explicit to prevent unexpected dependency drift. Explicit dependency management, including the practice of version pinning and locking files (e.g., Maven’s dependency-lock plugin), keeps the system consistent and helps in establishing reproducible builds.

\textbf{Outdated Dependencies:}
Unused dependencies, redundant features, components, files, and outdated documentation should be systematically removed to reduce maintenance overhead and potential security risks~\cite{Owasp}. Continuous dependency monitoring, along with the use of automated update tools such as Dependabot and Renovate, can help mitigate risks with outdated libraries. Furthermore, maintaining a disciplined update policy aids in the timely application of security patches. Enabling automated pull request notifications may enhance more frequent and consistent dependency upgrades~\cite{mirhosseini2017can}.

\textbf{Suspicious Comments:}
Developers should adopt clear coding standards discouraging inactive commented-out code and establish regular peer-review practices to remove ambiguous comments promptly. Furthermore, establishing clear guidelines on the types of information that should be included in code comments and enforcing these guidelines through code reviews can enhance coding standards~\cite{rahman2019seven}.

\textbf{Wildcard Usage:}
A wildcard dependency may be helpful in the early phase of software development, but it has downsides as the project progresses. While wildcards ensure that dependencies are always up-to-date, they also introduce risks by pulling in potentially unstable or breaking changes without proper review~\cite{socket}. Dependency versions should be explicitly specified or use a floating range. Employing lock files and version pinning would enhance reproducibility and secure builds by preventing uncontrolled upgrades.

\subsection{Implications of Findings}\label{implication}
Our findings have several practical implications. First, the prevalence of security-related smells (e.g., hardcoded secrets and insecure URLs) suggests a need for greater security awareness during build script development. Second, the presence of maintainability-related smells, such as missing version information and inconsistent dependency management, points to the need for improved tooling and standardization practices. Organizations can adopt our tool as part of code review to enforce best practices.
Moreover, the detection of smells across open-source repositories indicates that these issues are widespread, underlining the need for ecosystem-wide adoption of static analysis tools, such as Sniffer, to build systems for better maintainability and reduce technical debt.
\subsection{Future Work}~\label{future-work}
While our work lays the foundation for detecting code smells in build scripts, several directions remain for future research. To enhance the tool’s detection capabilities, techniques such as machine learning or natural language processing could be integrated to improve its accuracy, especially for context-sensitive or semantically subtle smells. Incorporating contextual information such as project maturity, domain, or deployment environment can help prioritize smells based on impact and relevance. Expanding the tool’s applicability to other build systems and configuration files, such as Bazel or Ant, would broaden its utility. In this work, we focused on zero-shot prompting to prioritize generalization and minimize computational cost, as our goal was to use LLMs filtering and exploratory support. In future work, the use of more refined prompting strategies (e.g., few-shot, Chain-of-Thought) may improve the LLMS detection performance. Additionally, incorporating developer feedback or crowdsourced labeling could refine detection accuracy and offer insights into how developers interact with smells. Finally, longitudinal and cross-project analyses may reveal patterns in how smells evolve over time and vary across domains.

\section{Ethical Consideration}\label{ethics}

This study involved the analysis of publicly available open-source build scripts collected from GitHub. No private or personally identifiable information (PII) was accessed or used. All data was handled in compliance with the platform's terms of service and policies~\cite{github:2023}. We also obtained institutional review board (IRB) approval for our user study. All data was collected, handled, and stored in an institutional secure storage. Our goal is to support the broader software engineering community in improving build quality and security. 

\section{Threats To Validity}\label{thtb}
In this section, we acknowledge and discuss the limitations of our research findings:
\\
\textbf{Conclusion Validity:}
The identification and classification of code smells in build scripts involved subjective judgment. The initial extraction, categorization of smells, and mapping to CWEs were performed manually by the authors, introducing potential subjectivity.
Different researchers might classify the same code differently based on their experience and perspectives. However, to overcome the obstacle, we followed a systematic way to validate our findings by multiple authors involved in the coding and resolved the disagreement by following established guidelines~\cite{campbell2013coding}.

\textbf{Internal Validity:}
We acknowledge the possibility of other code smells existing within build scripts that were not identified in our study. To mitigate this threat, we analyzed 2134 build scripts across Maven, Gradle, Make, and CMake; additional or context-specific smells may remain undiscovered. In the future, we aim to expand our research by expanding the dataset and exploring additional scripting contexts to enhance the comprehensiveness of identified smells.
The detection accuracy of the developed tool is dependent on heuristic-based rules and patterns defined during tool development. These heuristics could produce false positives and false negatives. To address this, we iteratively refined our heuristics through continuous testing against manually validated scripts.

\textbf{External Validity:}
Our findings are subject to limitations in external validity, as the results may not be generalizable beyond the studied build scripts. The tool is specifically developed for certain build script types, and thus, findings might not be directly extended to other build systems with different syntactic or semantic structures. Furthermore, the evaluation was conducted exclusively on open-source build scripts obtained from GitHub repositories.
 The prevalence and impact of the identified code smells may vary in proprietary or enterprise environments, where development practices, quality standards, and tooling ecosystems differ.

\section{Conclusion}\label{conclusion}
Code smells indicate some recurring coding patterns that occur frequently. Though it may not always have negative effects, a code smell should still be taken seriously because it could be a sign of future security and maintainability risk. In this study, we performed an analysis of code smells in build scripts through a mixed-methods approach combining qualitative issue analysis and large-scale static analysis. By analyzing 5,882 build scripts from Maven, Gradle, CMake, and Make across 4227 open-source GitHub repositories, we identified 13 categories of code smells, totaling 10,895 occurrences. Our findings highlight that certain smells, such as Insecure URLs, Hardcoded Paths/URLs, and Wildcard Usage, are particularly prevalent across specific build systems. Furthermore, the co-occurrence analysis revealed strong associations among specific smell pairs, suggesting the presence of systemic patterns in configuration practices.
These insights underscore the need for improved tooling and development practices to detect and address code smells in build automation scripts. To support this, we proposed mitigation strategies aimed at improving the quality, maintainability, and security of build processes in modern software engineering.

\section*{Acknowledgement}\label{ack}
This work was supported and funded by the National Science
Foundation Grant No. 2207008. Any opinions expressed in
this material are those of the author(s) and do not necessarily reflect the views of the National Science Foundation.





\bibliographystyle{IEEEtran} 
\bibliography{ref} 



\end{document}